\documentstyle[preprint,eqsecnum,aps,epsf]{revtex}
\newcommand{\half}{\frac{1}{2}}
\newcommand{\ha}{\textstyle\frac{1}{2}}
\newcommand{\nn}{\nonumber}
\begin{document}
\tightenlines
%\draft
\preprint{KYUSHU-HET-43}
\title{RPA for Light-Front Hamiltonian Field Theory}
\author{Koji Harada}
\address{Department of Physics\\
Kyushu University\\
Fukuoka 812-8581\\ Japan}
\date{\today}
\maketitle
\begin{abstract}
A self-consistent random phase approximation (RPA) is proposed
as an effective Hamiltonian method in Light-Front Field Theory (LFFT). 
We apply the general idea to the light-front massive Schwinger model 
to obtain a new bound state equation and solve it numerically.
\end{abstract}


\section{Introduction}

One of the most interesting aspects of Light-Front Field Theory 
(LFFT)\cite{reviews} is that it shares many features with non-relativistic 
field theory, due to the simple vacuum. It makes the Hamiltonian method 
sensible and thus LFFT becomes an attractive framework for bound state 
problems. To my knowledge, however, only very limited techniques 
which are common in non-relativistic many-body problems have been used in the
LF context, e.g., the Tamm-Dancoff approximation\cite{PHW}
and a boson expansion method\cite{bsnexp}. In order to fully exploit the
advantages of LFFT, it is desirable to consider the possibility of the 
application of other many-body techniques to LFFT.

In this paper, we consider a kind of random phase approximation (RPA) applied
to bound state equations. In the usual RPA, only one-loop vacuum polarization
diagrams (rings) are summed up\cite{RPA}.
Here we replace the vacuum polarization with the ``meson'' intermediate states
obtained by the bound state equation, as we did in a previous paper\cite{bh}. 
Thus our bound state equation is a self-consistent one.

The motivation for this approximation are two-fold: (1) The most popular
method of solving the bound state problems in LFFT is Hamiltonian 
diagonalization based on the Fock representation.  
In the crudest approximation, a meson is described as a $q\bar q$ 
valence state.
(In general, such a description is very good in LFFT.)
However it becomes rapidly more difficult to include higher Fock 
states. This difficulty leads to the idea of 
effective Hamiltonians\cite{eff,elena} 
which include the effects of higher Fock states. A problem in this
approach is that the effective Hamiltonians are obtained only perturbatively.
In particular, it has not been considered to include {\em infinitely} many
Fock states. (2) In a previous paper\cite{bh} on the theta vacuum 
in the massive Schwinger model, we have shown that the vacuum polarization
effects are very important for the low-energy physics. It is therefore 
natural to consider such effects in the calculations of meson states even
in the absence of theta. Another important point in the previous work is
that the vacuum polarization is approximately represented as intermediate
``meson'' states obtained by the (two-body) bound state equation. 

In order to make the idea as concrete as possible, we consider the
massive Schwinger model as an example. In the next section, we review
the perturbative Tamm-Dancoff transformation\cite{eff} and apply it
to the massive Schwinger model. In Sec.~3, we obtain the self-consistent
RPA bound state equation. In Sec.~4, we numerically solve the equation.
The final section is devoted to the conclusion and discussions.

\section{Perturbative Tamm-Dancoff transformation}

In order to make the bound state calculation tractable, it is necessary to
make the Hamiltonian more diagonal in particle number space. In a previous
paper\cite{eff}, we eliminate the particle-number-changing interactions 
perturbatively by a similarity transformation. 
Although it was originally proposed to resolve the sector-dependent 
counterterm problem in LF Tamm-Dancoff approximation\cite{PHW},
it is also useful in the massive Schwinger model, where no ultraviolet 
divergence occurs.

Consider the Hamiltonian $H$ of the form,
\begin{equation}
  H=H_0+\lambda W+ gV
\end{equation}
where $(H_0)_{mn}=\Omega_m\delta_{mn}$. The interaction $W$ does not change
the particle number while $V$ does. The parameters $\lambda$ and $g$ are to
organize the perturbation theory. We will put $\lambda=g=1$ at the end.

We eliminate the particle-number-changing interaction perturbatively by
the following similarity transformation,
\begin{equation}
  H_{eff}=e^{iR}He^{-iR}
\end{equation}
so that $H_{eff}$ is required to be diagonal in particle number. The operator
$R$ is expanded in a power series of $g$ and $\lambda$,
\begin{equation}
  R=gR_1+g^2R_2+g\lambda T_1+ \cdots.
\end{equation}

To the second order in $g$, we obtain the following effective Hamiltonian,
\begin{eqnarray}
  H_{eff}&=&H+g^2\Delta H, \\
  (\Delta H)_{mn}&=&\half\sum_k
  \left(\frac{1}{\Omega_m-\Omega_k}+\frac{1}{\Omega_n-\Omega_k}\right)
  V_{mk}V_{kn}.
\end{eqnarray}

Although this effective Hamiltonian is rather simple, 
it however gets very complicated beyond the second-order
in perturbation theory. Furthermore, it is perturbative by construction
and we do not have enough control over non-perturbative dynamics.
For example, we do not know how to treat the non-perturbative 
divergences which occur in the diagonalization, except
for the resummation by the use of the running coupling constant.

In the case of the massive Schwinger model, however, there are no ultraviolet
divergences and, as is seen in the next section, we can generalize these
interactions (at least partially) to include some non-perturbative effects.

For the massive Schwinger model, the effective interactions
$\Delta H$ in the $q\bar q$ sector may be diagrammatically represented 
in Fig.~\ref{effint}.

\section{A self-consistent RPA bound state equation for the massive Schwinger
model}

\begin{figure}

\centerline{\epsfbox{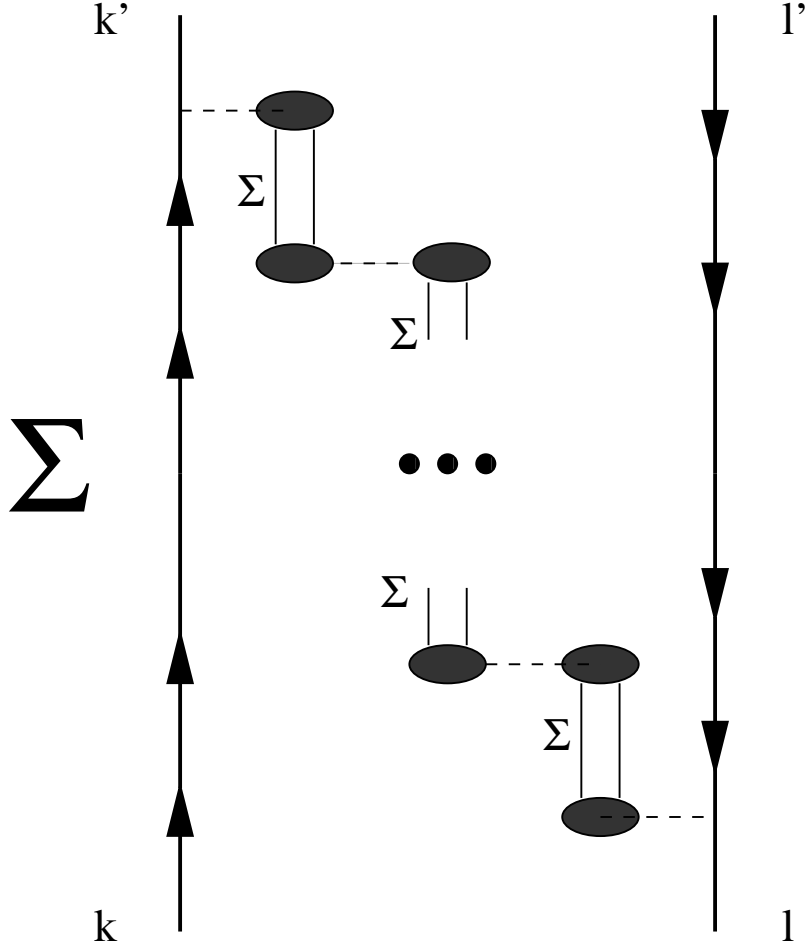}}

\vspace{0.5cm}

\caption{Effective Coulomb interactions with vacuum polarization being
included. The geometrical sum of the vacuum polarization as an approximate
complete set of ``meson'' states is taken. This figure is for $k'>k$ and 
$l>l'$.
There is a similar (North-East) diagram for $k>k'$ and $l'>l$.}
\label{eff_coulomb}
\end{figure}

Among these effective interactions, we are particularly interested in the
Coulomb interaction with vacuum polarization being included. 
In Ref.\cite{bh}, we showed that the vacuum polarization effects 
are very important in low-energy physics. 
It is therefore natural to include this effect
(i.e., charge shielding) in the bound state calculations.  Note that
this approximation is essentially equivalent to the so-called random
phase approximation (RPA) in many-body problems. Here, however, we do a bit
further. As we did in Ref.\cite{bh}, we replace the vacuum polarization
with an approximate complete set of meson states. 
A simple calculation gives the following replacement,
\begin{eqnarray}
  & &-(2\pi)e^2\sqrt{k'l'kl}\delta(k'+l'-k-l)\frac{1}{(k-k')^2} \nn \\
  &\rightarrow &
  -(2\pi)e^2\sqrt{k'l'kl}\delta(k'+l'-k-l)\frac{1}{(k-k')^2} \\
  & &{}\times\left\{1+\frac{e^2}{2\pi}\frac{1}{(k-k')^2}\sum_n|g_V(n)|^2
    \left(\frac{1}{\frac{m^2}{ll'}+\frac{\mu_n^2}{(l-l')^2}}
      +\frac{1}{\frac{m^2}{kk'}+\frac{\mu_n^2}{(k-k')^2}}\right)\right\}^{-1}, \nn
\end{eqnarray}
where $g_V(n)=\int_0^1 dx \psi_n(x)$ and $\psi_n(x)$ is the wave
function of the $n$-th ``meson'' with mass $\mu_n$. (See Ref.\cite{bh}
for the convention.)
$k$ and $l$ ($k'$ and $l'$) are longitudinal momenta of the 
incoming (outgoing) fermion and antifermion respectively. 
See Fig.~\ref{eff_coulomb}.
Note that we have summed the geometric series
of the vacuum polarization\cite{remark}.

With this replacement, we obtain the following bound state
equation (in units of $e/\sqrt{\pi}$),
\begin{equation}
  \mu^2\psi(x)=\frac{m^2}{x(1-x)}\psi(x)+\int_0^1 dx \psi(x) 
  +\int_0^1 dy \frac{(\psi(x)-\psi(y))}{(x-y)^2}
  S(\{\psi_n\},\{\mu_n\},m,x,y),
  \label{new}
\end{equation}
where
\begin{equation}
  S(\{\psi_n\},\{\mu_n\},m,x,y)=\left\{1+\half\sum_n|g_V(n)|^2 
    \left(\frac{1}{\mu_n^2+\frac{(x-y)^2m^2}{xy}}
      +\frac{1}{\mu_n^2+\frac{(x-y)^2m^2}{(1-x)(1-y)}}\right)\right\}^{-1}
\label{screeningfactor}
\end{equation}
is the screening factor. Eq.~(\ref{new}) is our extension of the 
't~Hooft-Bergknoff equation\cite{thooft:bergknoff},
\begin{equation}
  \mu^2\psi(x)=\frac{m^2}{x(1-x)}\psi(x)+\int_0^1 dx \psi(x) 
  +\int_0^1 dy \frac{(\psi(x)-\psi(y))}{(x-y)^2}.
  \label{old}
\end{equation}

Some remarks are in order:
\begin{enumerate}
\item The screening factor $S$ depends not only on the momentum transfer
$(x-y)$ but on the individual momenta of the fermion and the antifermion.
The $x$-dependent terms do not scale in a simple way.
It makes the analysis of this equation very complicated.

\item The equation (\ref{new}) is a self-consistent one, so that we should
solve it iteratively as we do in the next section. It is important to note 
that, although we started with the two-body bound state equation (\ref{old}),
our bound state equation (\ref{new}) contains the effects of 
{\em infinitely} many Fock states because of this self-consistency.

\item Although the equation (\ref{new}) is a self-consistent one, 
it is not the usual Hartree-Fock
approximation in which only the wave function of the specific state
is included. Our equation requires the information on {\em all\/} the
``meson'' states. It is not the usual RPA either, in which  only
the perturbative vacuum polarization diagrams (rings) are summed and 
the self-consistency is not required.

\item $0<S<1$. Of course
it physically means that the electric charge is screened 
due to the vacuum polarization. The effective charge reduces in a 
momentum dependent way.

\item The screening factor is included not only for the usual Coulomb term
but also the so-called ``self-induced inertia'' term,
\begin{equation}
  \left(\int_0^1 dy \frac{1}{(x-y)^2}\right)\psi(x).
\end{equation}
This is necessary 
in order to ensure the positivity of the eigenvalue $\mu^2$. 
By multiplying $\psi^\dagger(x)$
and integrating over $x$ from $0$ to $1$, we obtain
\begin{eqnarray}
  \mu^2\int_0^1 dx |\psi(x)|^2&=&m^2\int_0^1 dx\frac{ |\psi(x)|^2}{x(1-x)}
  +\left|\int_0^1 dx\psi(x)\right|^2 \nn \\
  & &{}+\half\int_0^1 dx\int_0^1 dy\frac{|\psi(x)-\psi(y)|^2}{(x-y)^2} 
  S(\{\psi_n\},\{\mu_n\},m,x,y).
\end{eqnarray}
Since the screening factor $S$ is positive, 
the eigenvalue $\mu^2$ must be positive. If we did not include the screening
factor in the self-induced inertia term, we would not have had the positivity
of $\mu^2$.

\item In the zero-momentum-transfer limit $(x-y)\rightarrow 0$, the screening
factor becomes momentum independent, i.e.,
  \begin{equation}
     S(\{\psi_n\},\{\mu_n\},m,x,y)\rightarrow 
     \left\{1+\sum_n\frac{|g_V(n)|^2}{\mu_n^2}\right\}^{-1}.
  \end{equation}
The factor $\sum_n|g_V(n)|^2/\mu_n^2$ is the one we encountered in 
Ref.\cite{bh}.

\item The wave function $\psi(x)$ for the 't~Hooft-Bergknoff equation 
(\ref{old}) has a definite meaning that it is the two-body amplitude 
with respect to the free fermion and antifermion. The wave function
$\psi(x)$ for our equation (\ref{new}) does not have such a clear 
meaning because it contains the contribution from the higher Fock
states. Roughly speaking, however, it could be considered as the
two-body amplitude with respect to the dressed (constituent) fermion 
and antifermion.
\end{enumerate}

\section{Numerical solution}
\label{numerical}
In this section, we numerically solve our bound state equation (\ref{new}).
Unfortunately, however, we are unable to find a useful set of
basis functions\cite{basis} which enable us to calculate the Hamiltonian 
matrix elements analytically.
Instead, we use a naive discretization of the equation. We do not claim
that the numerical results we present in this section is accurate. Rather
we emphasize that our equation gives a reasonably small correction
to the corresponding solution of the 't~Hooft-Bergknoff equation and 
therefore our equation could be a starting point of further improvements.

The method is extremely simple. We discretize the range of $x$, $[0,1]$,
into $N+1$ pieces, and impose the boundary condition on the wave function,
$\psi(0)=\psi(1)=0$. Thus our equation becomes an $N\times N$ matrix 
eigenvalue equation. We first solve it with $S\equiv 1$, and then use the
result to calculate the screening factor for the next calculation.
In this way, we iteratively solve the equation until the solution becomes
self-consistent. 

Most of the calculations were done with $N=2000$. 
We stop the iteration when the difference of the 
subsequent wavefunctions becomes less than $\epsilon=1.0\times10^{-12}$,
\begin{equation}
  \int_0^1 dx \left|\psi_{new}(x)-\psi_{old}(x)\right|^2 <\epsilon.
\end{equation}
A typical number of iterations is 4. In general, the convergence is slower
for smaller fermion masses.

\begin{figure}

\centerline{\epsfbox{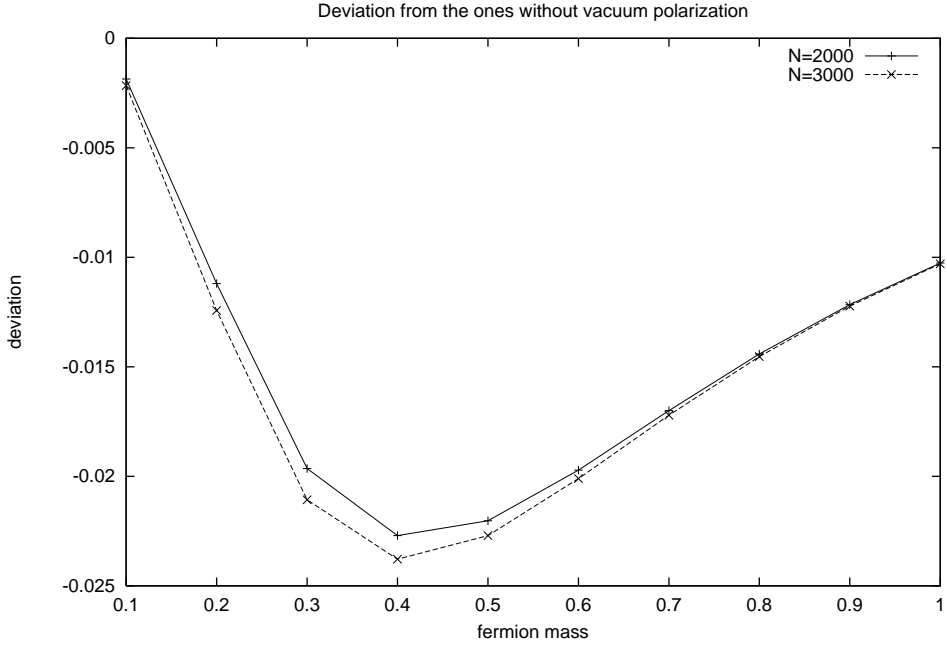}}

\vspace{0.5cm}

\caption{The deviation of the meson mass with the new bound state equation 
from that without vacuum polarization. Data are for $N=2000$ and 
for $N=3000$. The deviation is almost independent of $N$.}

\label{deviation}
\end{figure}

\begin{figure}

\centerline{\epsfbox{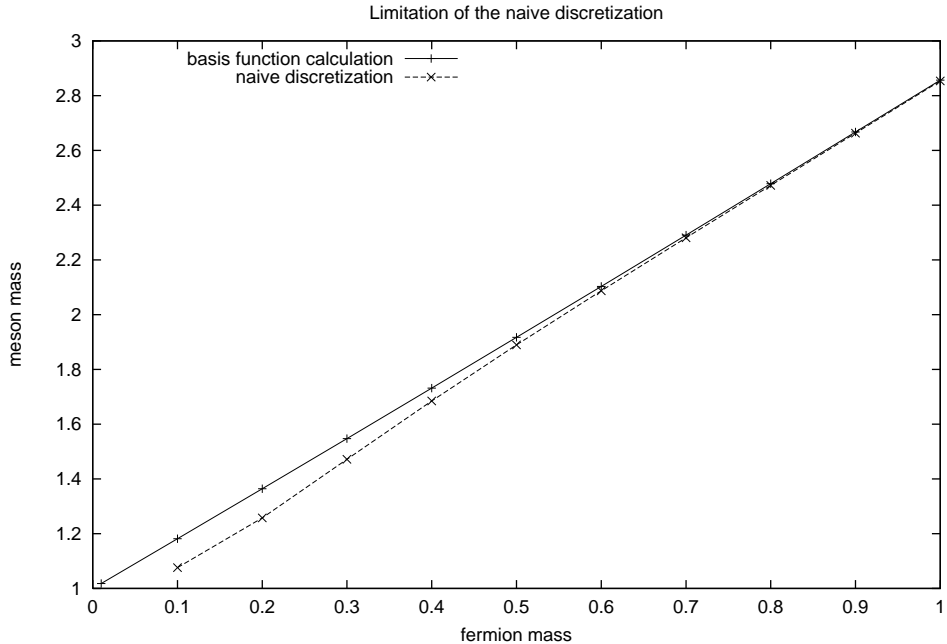}}

\vspace{0.5cm}

\caption{The limitation of the naive discretization. The results with the 
naive discretization are compared with those obtained by basis function 
calculation for the 't~Hooft-Bergknoff equation, the latter being considered
to be accurate for small fermion masses. The naive discretization does not 
produce accurate values for small fermion masses. }
\label{limitation}
\end{figure}

First we compare the results with those without vacuum polarization. 
Fig.~\ref{deviation}
shows how the meson mass calculated iteratively with the new bound state 
equation deviates from the one obtained with the 't~Hooft-Bergknoff equation.
We see that there is systematic improvement, which is almost 
independent of how fine 
the equations are discretized. For small fermion masses $m \alt 0.2$, 
because of the singular behavior of the wavefunction at the edges,
even the result without vacuum polarization depends on $N$, and therefore
the results there should not be trusted. See Fig.~\ref{limitation} for the
copmarison with the result obtained by the basis function method\cite{basis}
which is accurate for small fermion masses.
We however emphasize that the modification
due to vacuum polarization does not ruin the spectrum but leads to a 
reasonable correction. This is not so trivial because the modification is
essentially nonlinear and depends all the eigenvalues as well as 
the eigenstates. This, together with the valence dominance in LF massive 
Schwinger model\cite{valence}, gives a consistent picture that the 
modification with vacuum polarization correctly reflects the (small) 
effects of higher Fock states.

\begin{figure}

\centerline{\epsfbox{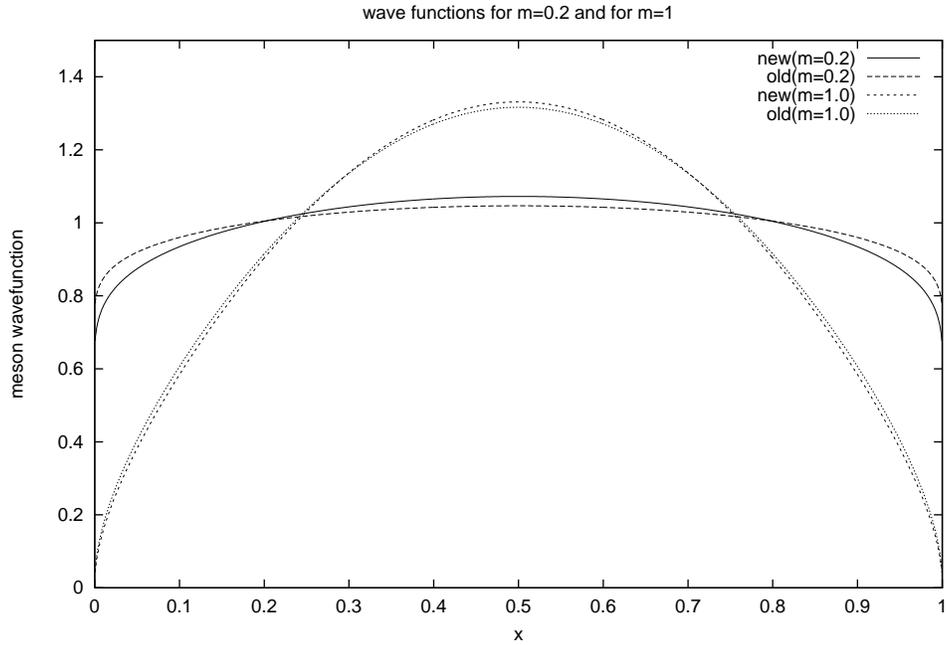}}

\vspace{0.5cm}

\caption{The comparison of the wavefunctions obtained for 
the (``old'') 't~Hooft-Bergknoff equation  and those for 
the ``new'' equation for $m=0.2$ and $m=1.0$.
The wavefunctions for the new equation are less singular at the edges
for small fermion masses.}
\label{comparewf}
\end{figure}

Next we compare the wavefunctions. As seen in Fig.~\ref{comparewf}, the
meson wavefunction for the new equation is less singular at the edges
($x\sim 0$ and $x\sim 1$). 

As we emphasized earlier\cite{remark}, 
apparently we should not sum up
the geometrical series of vacuum polarization. Here we did it as a 
regularization. In the calculations without summing up the geometrical
series, we numerically found that a state whose wavefunction is {\em odd}
under $x \leftrightarrow 1-x$ becomes the lightest one when the fermion
mass is small. This seems to be because 
\begin{equation}
  1-\half\sum_n|g_V(n)|^2\left(\frac{1}{\mu_n^2+\frac{(x-y)^2m^2}{xy}}
      +\frac{1}{\mu_n^2+\frac{(x-y)^2m^2}{(1-x)(1-y)}}\right)
\end{equation}
becomes very small when $m\rightarrow 0$. We are unable to give any
positive interpretation of such a lightest state with wavefunction
being odd under  $x \leftrightarrow 1-x$, and thus think that it is
an artefact. In fact, the odd lightest state occurs for $m \alt 0.1$, 
where the results should not be trusted because, as we remarked earlier,
the discretization cannot 
represent the correct endpoint behavior of the wave function, which is
very important for the correct spectrum.

We emphasized that the new bound state 
equation contains the information on all the ``meson'' states. It is 
however dominated by the lowest meson state and effectively reduces to 
a much simpler one. Namely, we numerically find that the sum in the 
screening factor (\ref{screeningfactor}) is strongly dominated by 
the first term. This fact reduces the CPU time considerably. 
In Table \ref{comparison}, we compare the reduced calculation with the
full one.

\section{Summary and discussions}
\label{summary}

In this paper, we developed a self-consistent RPA 
as a new effective Hamiltonian method in LFFT. The method was applied to 
the massive Schwinger model and a generalization of the 't~Hooft-Bergknoff
equation was obtained, which effectively includes the contributions 
from {\em infinitely} many Fock states.
We numerically solved it and found that the modification, i.e., the 
inclusion of vacuum polarization, leads to reasonable small corrections.
We hope that this new bound state equation becomes a starting point of
more elaborate approximations.

In the following, we would like to discuss several aspects about
the new bound state equation.

It has been shown that the first-oder coefficient $M_1$
of the expansion of the meson mass squared in terms of the fermion mass,
\begin{equation}
  \mu^2(m)=1+M_1 m +M_2 m^2 + \cdots,
\end{equation}
cannot be improved even if the variational space is systematically 
extended\cite{hhs}.
Later, Dalley suggested that the endpoint behavior of the wave function
may change if the effects of higher Fock states correctly. (It has been
assumed that the endpoint behavior is solely determined by the two-body
sector\cite{basis,hhs}.) Because the coefficient is closely related to the 
endpoint behavior, it is hoped that the so-called ``2\% discrepancy'' 
will disappear when the correct endpoint behavior is taken into account.
Since our new bound state equation contains the effects of higher Fock
states, it is natural to ask if the Dalley's suggestion is correct
quantitatively. For small fermion masses, it looks legitimate to approximate
the screening factor in the following way,
\begin{equation}
  S(\{\psi_n\},\{\mu_n\},m,x,y)\rightarrow {\cal S}\equiv
  \left\{1+\frac{|g_V(1)|^2}{\mu_1^2}\right\}^{-1},
  \label{slimit}
\end{equation}
where $\mu_1=\mu$ is the meson mass and $g_V(1)=\int_0^1dx\psi(x)$, 
and the bound state equation becomes
\begin{equation}
  \mu^2\psi(x)=\frac{m^2}{x(1-x)}\psi(x)+\int_0^1 dx \psi(x) 
  + {\cal S}
  \int_0^1 dy \frac{(\psi(x)-\psi(y))}{(x-y)^2}.
\end{equation}
We estimate the coefficient $M_1$ by using the ansatz 
\begin{equation}
  \psi(x)=x^\beta (1-x)^\beta,
\end{equation}
and ${\cal S}=\ha(1+{\cal O}(m))$, and find that $M_1=2\pi/\sqrt{6}=2.56510$
which should be compared with the value for the 't~Hooft-Bergknoff equation,
$2\pi/\sqrt{3}=3.62760$ and the `correct' value, $2e^{\gamma}=3.56125$.
Although it is considerably too small, it is important that it shifts from the
value of the 't~Hooft-Bergknoff equation. 
It is conceivable that the approximation employed here is too naive and
a more elaborate calculation would give a much better value.
In fact, the above approximation (the limit (\ref{slimit})) 
is NOT legitimate, because for finite $m$, however tiny, 
the term with the fermion mass is not negligible for $x< 1/m^2$. 
At present, we are unable to extract
any useful information on the endpoint behavior because of this subtlety.
The numerical results in the previous section may suggest that
the effects of higher Fock states would solve the 2\% discrepancy, though
the present numerical method is too crude to correctly calculate 
the meson mass in the small fermion mass region.

\begin{figure}

\centerline{\epsfbox{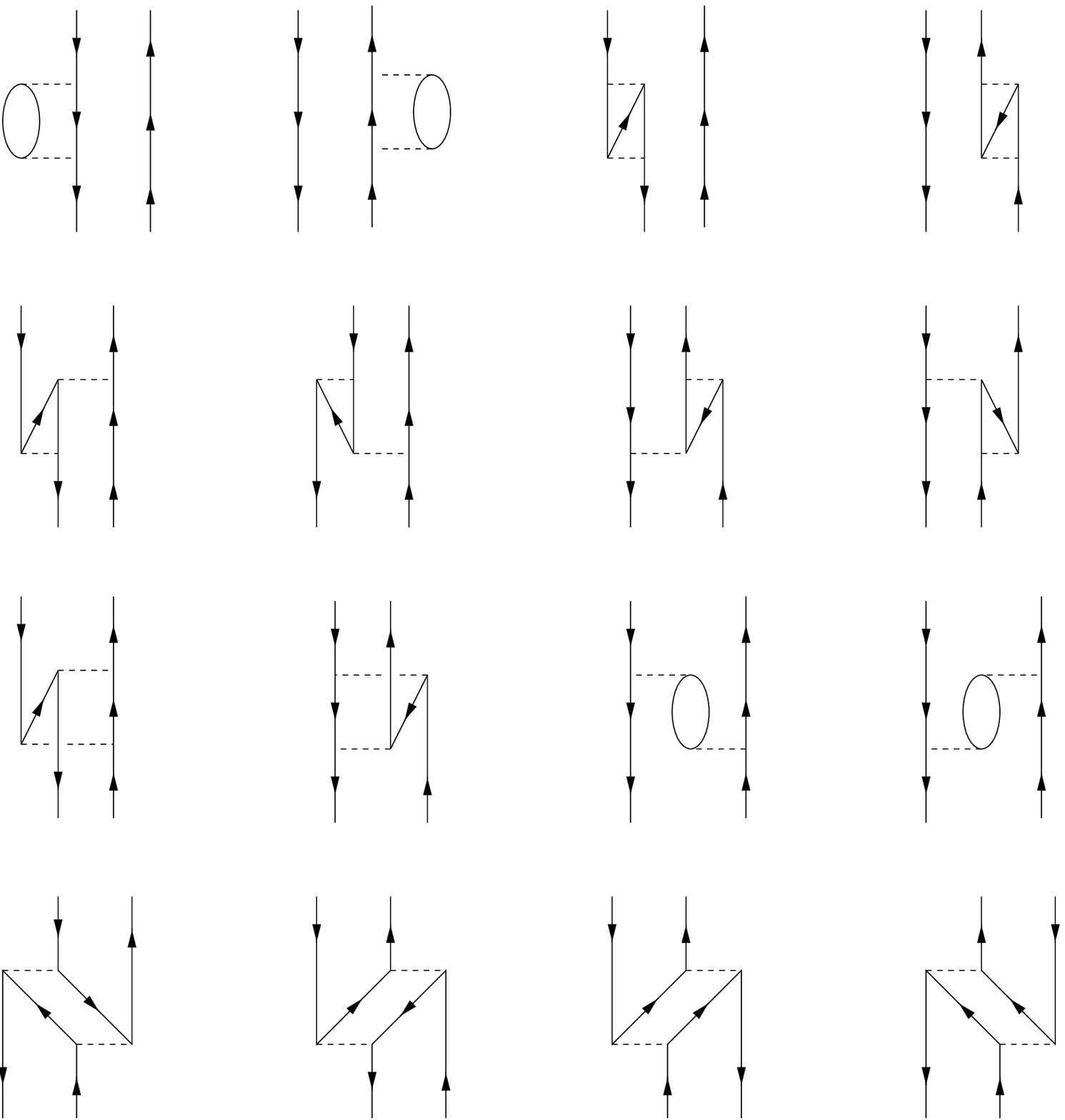}}

\vspace{0.5cm}

\caption{Effective interactions in the massive Schwinger model in the
$q\bar q$ sector, generated by a perturbative Tamm-Dancoff 
transformation to order $g^2$. Horizontal dashed lines stand for the
instantaneous Coulomb interactions.}
\label{effint}
\end{figure}

Although we included the effects of vacuum polarization as an effective
interaction, from the similarity transformation point of view, there are 
other kind of effective interactions. To ensure the invariance of the 
eigenvalues (``meson'' masses) under the addition of effective interactions,
it is necessary for the transformation to be unitary. That is, we have to
add other interactions in Fig.~\ref{effint} too. In this paper, we assumed
that the effects of the other interactions are small. But this is a cheat:
the self-energy diagram with vacuum polarization is infrared divergent.
Presumably, such infrared divergences cancel in the calculation of the
meson spectrum. (Some authors also assumed the cancellation\cite{Brasil}.)
We are however unable to pin point how such cancellation
occurs. Note that the cancellation of infrared divergences would be 
different from that in scattering processes. In QED, for example, 
because positronium is a neutral particle and has a finite size, it cannot
couple to very soft photons, which play an important role in the cancellation
of infrared divergences in scattering processes.

Finally, we give some comments on the outlook.
\begin{enumerate}
\item It is of course interesting to extend the present formulation to
realistic four-dimensional models. We are now working on this direction.

\item It is interesting to note again that the new equation is a 
self-consistent one from the spontaneous-symmetry-breaking point of view.
We hope that a similar setting would leads to a useful framework for the
vacuum problem in LFFT.

\item The vacuum polarization inside the meson is also important in the
presence of theta. It solves several puzzles concerning theta.
The work is now in progress.
\end{enumerate}

\acknowledgements

The author would like to thank T.~Heinzl for the discussions at the very
early stage of this investigation. He is grateful to Y.~Yamamoto,
M.~Taniguchi, and K.~Inoue for the discussions, especially on the IR
property, to O.~Abe for the discussion on the 2\% discrepancy and his
work, and to Y.~R.~Shimizu, S.-I.~Ohtsubo, and K.~Itakura, for
the discussions mainly on the many-body techniques. He is also 
grateful to M.~Burkardt for comments and for the discussions on the case
with theta.

%%%%%%%%%%%%%%%%%%%%%%%%%%%%%%%%%%%%%

%%%%%%%%%%%%%%%%%%%%%%%%%%%%%%%%%%%%%%%%%%

%%%%%%%%%%%%%%%%%

%%%%%%%%%%%%%%%%%

%%%%%%%%%%%%%%%%%

%%%%%%%%%%%%%%%%%

%%%%%%%%%%%%%%%%%

\begin{table}
  \caption{Comparison of the reduced calculation with the full one. 
The mass of the lowest meson state is shown. In the reduced calculation, 
we take only the first term in the sum in the screening
factor (\ref{screeningfactor}). The calculations are done with the naive
discretization explained in Sec.~\ref{numerical} with $N=2000$.}

\label{comparison}

\vspace{0.5cm}

\begin{tabular}{|c|c|c|} 
   fermion mass & reduced calculation & full calculation \\ \tableline
   0.1&1.07407458&1.07407437 \\
   0.5&1.84809454&1.84775659 \\
   1.0&2.82537904&1.84775659 \\
   5.0&1.06492610&1.06485924 \\ 
 \end{tabular}
\end{table}


\begin{references}
\bibitem{remark} Perhaps we should not sum the geometric series
because the meson states apparently contain such contributions in themselves. 
In Ref.\cite{bh}, we did not sum the series. See Sec.~\ref{summary} for the
discussion.
\end{references}
\end{document}